# Reception Reader: Exploring Text Reuse in Early Modern British Publications

**DAVID ROSSON**
**EETU MÄKELÄ**
**VILLE VAARA**
**ANANTH MAHADEVAN**
**YANN RYAN**
**MIKKO TOLONEN**

*Author affiliations can be found in the back matter of this article

## ABSTRACT

The Reception Reader is a web tool for studying text reuse in the Early English Books Online (EEBO-TCP) and Eighteenth Century Collections Online (ECCO) data. Users can: 1) explore a visual overview of the reception of a work, or its incoming connections, across time based on shared text segments, 2) interactively survey the details of connected documents, and 3) examine the context of reused text for "close reading". We show examples of how the tool streamlines research and exploration tasks, and discuss the utility and limitations of the user interface along with its current data sources.

# (1) INTRODUCTION

The Reception Reader[1] is a tool designed to help humanities scholars study the reception of works over time. Users can search for a work of interest in the Early English Books Online (EEBO-TCP)[2] and Eighteenth Century Collections Online (ECCO)[3] databases which cover a considerable portion of books published in Britain between 1450 and 1800. With ease, users can navigate through instances of text reuse organised temporally and clustered around different sections of the document. The design of the tool optimises for the use case of switching between distant and close reading for a deeper understanding of reception and influence.

The study of context and reception is critical for understanding early modern intellectual, literary, and cultural history (Skinner, 1969; Thompson, 1993; Jauss, 2019). Historically, this was done through manual methods, but with the advent of digital humanities, there has been a growing interest in systematic and data-driven studies of both reception and intertextuality between works (*e.g.* de Bolla et al., 2020; Gavin, 2016; Ladd, 2021).

Text reuse detection using computational methods has become a common practice in digital humanities (Gladstone & Cooney, 2020; Vesanto et al., 2017; Salmi et al., 2020; Büchler et al., 2014; Citron & Ginsparg, 2015; Cordell, 2015; Lee, 2007; Mullen, 2016; Smith et al., 2013) and makes it possible to cover a vast number of works and a greater extent of shared passages despite noisy data. While text reuse does not necessarily imply influence, it provides valuable information for understanding reception at scale.

The Reception Reader is intended for scholars and students in the humanities, such as literature, history, and philosophy, and serves as a test case for the integration of technology and humanities scholarship in research. The broader goal is to empower humanists and bring computational research to a wider audience through researcher-focused tools and easier access to data. The aim is not only to provide a tool for research, but also to promote data literacy. The Reception Reader can be used in educational settings, such as project courses, to introduce humanities students to the world of data science through hands-on experience and give them a taste of what is possible in the field of digital humanities.

# (2) THE USER INTERFACE

This section serves as an introductory guide for using the Reception Reader.

## (2.1) METADATA SEARCH

Start with the search page to look for a document of interest. The query text can be an author's name, one or multiple words in the book's title, or some combination of these. Prefix and fuzzy searches are supported. Click the search button or hit the "Enter" key to continue. The results are scored based on how many terms are matched, and the first 100 results are returned, sorted by the match score.

The document ID, publication year, author name, and short title of the document are shown in the search results (Figure 1). Click on a column's header to sort that column and toggle between ascending and descending order. Secondary sort order carries over on re-sorting, for example, if you first click on sort by year, then click on sort by author, the results will be sorted by author name, while rows with the same author name are sorted by year. You can also use the browser's text search to navigate around the results. Clicking on the document ID of a result will select it as the primary document and open up the main interactive interface.

## (2.2) INTERACTIVE VISUALISATION

The main interface has two sections: on the left is a beeswarm chart of the reuse connections, on the right is a view of the reuse context.

In the beeswarm chart, each dot represents an instance of reuse connected to the primary document. Documents with the same author name are excluded from the results. By default, the "outgoing" connections are displayed, *i.e.* documents with a publication year equal to or later than that of the primary document. To see incoming connections instead, click on the "In" button in the toggle group above the top right corner of the chart. A set of time range filters can be used to show only dots between certain years. Enter a year number in the input box for "from" or "to" or use the up and down arrow keys to adjust the year number, then hit "Enter" to apply the filter.

---

1   https://www.receptionreader.com.

2   https://textcreationpartnership.org/tcp-texts/eebo-tcp-early-english-books-online, accessed 3 April 2023.

3   https://www.gale.com/intl/primary-sources/eighteenth-century-collections-online, accessed 3 April 2023.

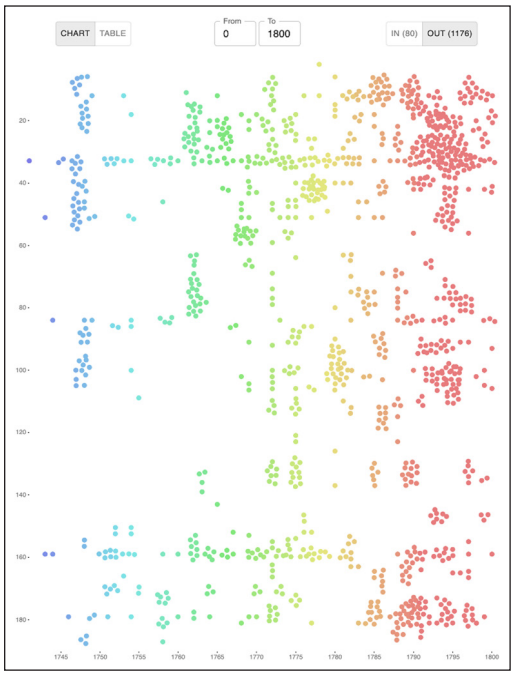

**Figure 1** Searching for documents by metadata.

The dots are placed along the X-axis by publication year, and along the Y-axis by linear location of the reused text in the primary document (axis labels show page numbers). A dot's colour maps to the gap between publication years. In the "outgoing" view, blue dots were published shortly after the primary document and red ones half a century or more later.

**Figure 2** Visualisation of outgoing edges.

The visualisation reflects, to some extent, which parts of the document are most reused. In a beeswarm chart, dots are first placed by their X and Y values then go through a force-directed Voronoi simulation with equal weighting across the dots (Figure 2). The position of the dots is then "nudged" in multiple rounds so they cluster around the original position but avoid overlapping on top of each other, the way individual bees in a swarm avoid collision. This charting mechanism can be leveraged to visualise reception in terms of clustering patterns. For example, sections that have been reused constantly throughout the period will have dots line up horizontally around a certain page range (Y value). Vertical bands can be used to spot large-scale copying throughout the document at different times. A side-way "cone" that broadens (in height) over time suggests growing reuse of that part of the document.

Interact with the chart by hovering over a dot. A tooltip shows the publication year, author, and title of the connected document and the page number of the reused segment in the primary document. Clicking on a dot shows the context of the reused segment in the page viewer area on the right-hand side of the screen.

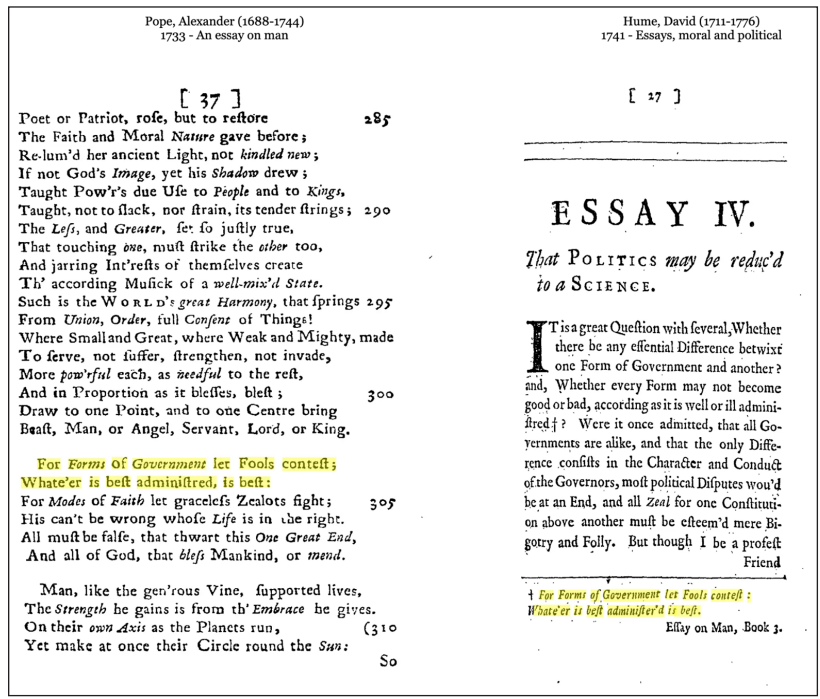

**Figure 3** Viewing the context of text reuse segments.

A scanned image of the page where the reuse is found, when available, is displayed, with the reused text highlighted, for both the original document and the connected document (Figure 3). Reading the context surrounding the reused text enables the examination of the primary evidence of discourse and reception: what was reused, how it was reused, and in many cases, further commentary by the later author on the quoted passage and the earlier author.

**Figure 4** Table view of the edges connected to the primary document.

| | From | To | | |
|---|---|---|---|---|
| CHART TABLE | 1741 | 1800 | | IN (80) OUT (1176) |

| ↑ Year | Author | Title | Page |
|---|---|---|---|
| 1741 | Merchant | An essay on the improvement of the woollen manufacture | 33 |
| 1743 | Bolingbroke, Henry St John (1678-1751) | A dissertation upon parties | 51 |
| 1743 | Pope, Alexander (1688-1744) | The works of Alexander Pope, Esq | 159 |
| 1744 | Shaftesbury, Anthony Ashley Cooper (1671-1713) | Characteristicks | 84 |
| 1744 | Pope, Alexander (1688-1744) | Epistles to several persons | 159 |
| 1745 | Pope, Alexander (1688-1744) | An essay on man | 33 |

The list of connected documents can also be displayed in a table view with sortable columns and browser-native text search (Figure 4). Each row corresponds to the information of a dot in the chart view. Clicking on a row will load the reuse context in the page viewer.

## (2.3) TECHNICAL SETUP

In earlier days of the broader project, the data points of the reuse edges were pre-processed and stored in a number of Parquet files which were then queried using a local Spark cluster. The Parquet files were not indexed and the queries were not optimised which resulted in each query taking a few minutes. For end-user facing applications, we needed the queries to be returned within "online" response times. We set up a MariaDB service and ported the edges data, available both as column-stores and as indexed tables, queryable through SQL. Reused text snippets were omitted so that the tables would be of a manageable size.

| t1_id | t1_start | t1_end | t2_id | t2_start | t2_end | align_length | positives_percent |
|---|---|---|---|---|---|---|---|
| 0 | 74,995 | 77,077 | 1 | 128,157 | 130,191 | 1,549 | 84.64 |
| 0 | 73,223 | 74,988 | 2 | 123,992 | 125,772 | 1,365 | 87.62 |
| 0 | 73,249 | 74,988 | 3 | 41,528 | 43,307 | 1,360 | 85.51 |
| 0 | 75,145 | 77,074 | 4 | 166 | 2,177 | 1,494 | 78.85 |
| 0 | 100,832 | 101,960 | 5 | 1,631,767 | 1,632,851 | 832 | 91.47 |
| 0 | 100,832 | 101,964 | 6 | 1,587,533 | 1,588,637 | 837 | 90.92 |
| 0 | 100,825 | 101,960 | 7 | 262,080 | 263,198 | 840 | 88.21 |
| 0 | 100,817 | 101,960 | 8 | 1,607,229 | 1,608,339 | 848 | 87.26 |
| 0 | 100,806 | 101,964 | 9 | 9,179 | 10,309 | 868 | 81.91 |
| 0 | 75,996 | 77,077 | 10 | 139,012 | 140,055 | 786 | 80.79 |
| 0 | 75,146 | 75,762 | 10 | 138,394 | 138,989 | 462 | 88.96 |
| 0 | 75,996 | 77,074 | 11 | 51,778 | 52,828 | 781 | 81.43 |
| 0 | 75,145 | 75,762 | 11 | 51,147 | 51,755 | 462 | 85.93 |
| 0 | 75,996 | 77,074 | 12 | 48,699 | 49,752 | 789 | 82 |
| 0 | 75,144 | 75,762 | 12 | 48,070 | 48,675 | 465 | 88.82 |
| 0 | 134,947 | 135,856 | 13 | 84,798 | 85,702 | 698 | 84.1 |

**Figure 5** Rows of reuse edges in an SQL database.

For the Reception Reader, we can query all the reuse instances connected to a primary document, and get the IDs of the connected documents and the location (start and end offsets) of the reused segment in each of the two documents respectively (Figure 5). A Node.js service was set up to receive requests from the web client, retrieve the rows of reuse instances from MariaDB, enrich the results with document metadata, and return them for presentation.

The location of a text segment within a document is specified by the character offset, which is the sequential index when the entire text of a document is treated as a string. The Octavo database (Mäkelä et al., 2020) provides page mapping information at the token level for ECCO documents, that is, for each word in the text, it returns a corresponding page number and the pixel coordinates of the word's image region on the scanned page. This information is used to show the page of the reuse context and text highlights. The offsets in Octavo are shifted from the offsets in the reuse data because the document text in Octavo contains extra annotations of chapter headings. Thus, Reception Reader's backend service also translates the offsets between the reuse data and the page mapping data.

The frontend (user interface) and backend (data services) are each set up as a Docker image and deployed on Rahti, the container cloud provided by CSC's computing infrastructure.

## (3) RECEPTION READER IN USE

### (3.1) WHAT CAN FIRST-TIME USERS DO WITH THIS TOOL?

Begin by searching for a familiar work or an author that is interesting to you and get an overall impression of the results. Select a document, open the chart view, and look for patterns in the reuse instances. Hover over the dots to see what documents are connected to the work of your interest, click on a dot to load the highlighted pages. Examine the context by reading around the reuse.[4] Then continue your exploration by searching for another work that is familiar to you or one that caught your attention in your previous search. This will help you get a sense of the idea of switching between distant and close reading and the versatility of the Reception Reader and how its use can be tailored to meet your specific research needs.

### (3.2) TYPES OF REUSE AND RESEARCH QUESTIONS

The Reception Reader is a resource for finding connections between works and exploring how various editions and works were reused in early modern British printed books. For instance, a researcher interested in the reception of Lord Shaftesbury's *A Letter Concerning Enthusiasm* (1708) could easily search for and select a relevant edition, and see all the reuses of it in seconds. If the researcher wishes to examine how Mary Astell's *An Enquiry After Wit* (1722) relates to *A Letter Concerning Enthusiasm*, the Reception Reader is the perfect tool for further investigating that pair's connection and context (Figure 6).

The instances of textual overlaps are numerous and of diverse types, ranging from quotes to reprints to publication artefacts. Specific approaches to identifying, refining, and analysing the relevant instances will largely depend on the research questions to be addressed.

---

4    If you are looking to conduct a detailed study of a particular work, the Reception Reader also links to Gale's interface for the collection, where you can access the complete text of each edition, given that your institution has a licence for it.

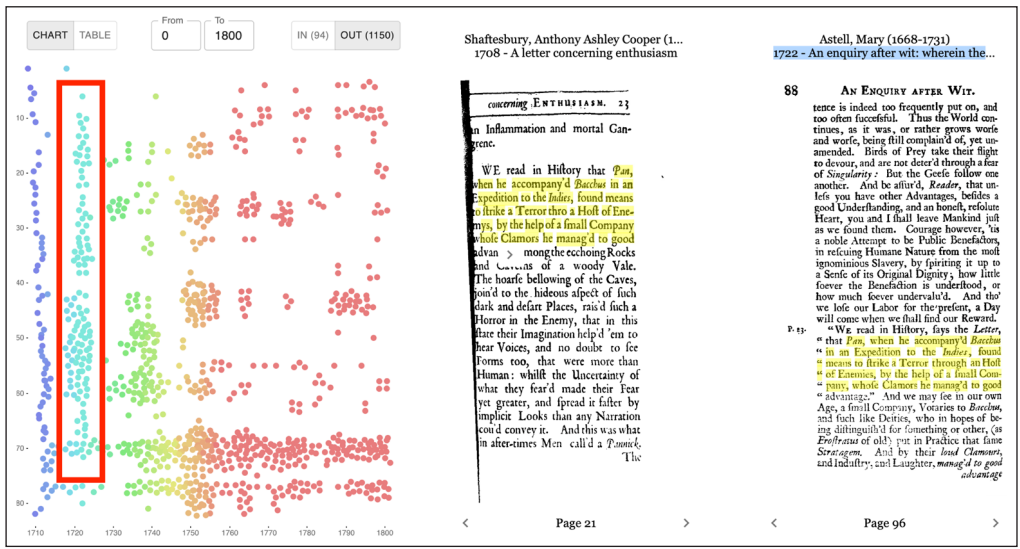

**Figure 6** Viewing reuses of Shaftesbury's *A Letter Concerning Enthusiasm*.

Overlaps with Astell's *An Enquiry After Wit* highlighted by a red box.

| TYPE OF REUSE | EXAMPLES | POSSIBLE RESEARCH QUESTIONS |
| --- | --- | --- |
| Quotes | Latin, biblical, famous quotes. | What was the process of quotes from Lucretius becoming epigraphs over time? |
| Reprints of longer passages | Reused sections or fragments from essays or treatises appearing in works by different authors. | What was the distribution of Hume's essays outside of his published works? |
| Modified reuse | Modified reuse of a specific work in another work. | How did Clarendon's *History of the Rebellion* feature in other historical works? |
| Verse reprints | Reprinting of poetry in unexpected or uncommon locations. | How did Dryden's poetry spread outside of known collections? |
| Unattributed reuse | Hidden or obscured reuse of texts. | Can we gain a broader understanding of the reception of Hume's essays by exploring their use in other works without proper attribution? |
| Artefacts | Imprint of publisher, advertisement. | What was the distribution pattern of advertising for Hume's *Treatise* in printed books in the eighteenth century? |

**Table 1** Examples of types of reuse that can be analysed.

For researching the reception history of a specific work and exploring unknown links, the Reception Reader offers a convenient way to study intertextuality in early modern British printed books by helping the user investigate both the incorporation of other texts into an author's work and the reuse of an author's work by others. The tool enables easy identification of the most frequently reused parts of an edition and the study of different levels of text reuse, including full reprints, partial and modified reprints, and quotes (Table 1).

The Reception Reader is particularly helpful in the study of the reception of a chosen edition, and as a tool to examine patterns of text reuse by a particular author. In a previous study, it was used to access all instances of text reuse of David Hume's *Essays* and gain new insights into their reception across the eighteenth century (Spencer & Tolonen, in press). In this case, the use of the tool was supplemented by the results of other research work, namely the use of chapter heading information to segment the volume of *Essays* into constituent parts.

In another case, the Reception Reader was used to examine the text reuse in Bernard Mandeville's works with the aim of discovering authorship patterns. This analysis showed that Mandeville's works reused parts from other authors, including Pierre Bayle, with a consistent form except in one work, *Modest Defence of Publick Stews*, suggesting that it may have been written by another author—a theory which has been put forward by other scholars in the field. Text reuse analysis may help uncover obscure evidence to solve or corroborate on such literary puzzles, and Reception Reader can help speed up the exploration.

# (4) DISCUSSION

## (4.1) DATA

The concept of "reuse" can be broad, ranging from allusive references (Manjavacas et al., 2019) to textual overlaps, with the latter narrower sense being the focus of this research tool and its datasets. Identifying textual overlaps in digitised sources is not trivial due to Optical Character Recognition (OCR) errors. The use of a modified version of BLAST (Vesanto, 2019)—a bioinformatics algorithm for finding regions of similar sequences, first experimented with Finnish newspapers (Salmi et al., 2020)—was shown to be an OCR error-resilient way of detecting textual overlaps. As part of the High-Performance Computing and Historical Discourses (HPC-HD) project, this technique was applied to EEBO-TCP and ECCO, namely, corpora of books printed in Britain between 1450–1800 (Tolonen et al., 2022; Lahti et al., 2019).

The search process was computationally intensive, and required the use of cluster computing resources provided by CSC (Finland) and the LUMI consortium. Compared to text reuse in newspapers, reused sequences in ECCO and EEBO-TCP, which contain predominantly books, are longer, and the collections are larger than newspaper collections. Thus, a minor reduction in detection quality was also needed to enable the computation to complete with reasonable resource usage. This process has led to a massive dataset of 961 million data points that provides the basis of text reuse data for the Reception Reader.

Comparing the results with a critical edition of *Enquiry Concerning Human Understanding* (EHU; Hume, 2000), we found that the BLAST-based method identified all instances of text reuse of Hume's *Treatise of Human Nature* included in EHU except those in footnotes. In the course of preparing for an article on the reception of Hume's essays (Spencer & Tolonen, in press), a processed list of 1,400 cases of text reuse derived from the BLAST results were manually checked, of which approximately 1,200 were confirmed to be in line with subject expert assessment, with the rest being various repetitions.

When looking at a list of text reuse connected to a document of interest, a historian might be concerned with the list being "complete", that it hasn't missed or left out certain instances. The BLAST-based technique gives us a higher level of assurance than before that we have detected a reasonably exhaustive extent of the textual overlaps given the noisy OCR text. The other side of this concern is that when such a list is very long, we want to make sure that it is not longer than necessary, namely, the detected instances should be true positives. One issue we encountered was that some larger sequences of reuse were detected as multiple smaller sequences, due to significant noise in the OCR text, that is, the true instances were broken up at noisy points and became fragmented. Another issue was that the pair-wise comparison of similarity does not account for the direction of reuse in time and generates many "duplicate" results when a piece of text is frequently reused across many documents. If hundreds of documents quote a biblical verse, instead of reflecting the tree graph of one source fanning out to many destinations, each instance is detected as being linked to every other instance, which inflates the number of edges greatly. The challenges of merging these clusters are discussed in depth in Vesanto (2019), but the problems of sliding windows and mismatched fragment sizes are vastly amplified when dealing with books compared to newspaper data.

Evaluating the accuracy of the BLAST-sourced results is a larger topic on its own, while this article focuses on introducing a research tool for explorative interaction with datasets that are similar in structure. Our research group will have an upcoming paper to describe and discuss the data and computation procedure in more detail, and address issues that emerged when adapting a technique originally applied to newspapers to datasets that include books, newspapers, and pamphlets.

## (4.2) DATA SAMPLE: SHAKESPEARE

We are releasing a subset of raw data, including all detected text reuse fragments and relevant metadata in EEBO-TCP and ECCO, related to the works of William Shakespeare. This sample dataset is a portion of the derived text reuse data and does not include surrounding OCR texts or images from ECCO. It is free to use for research with citation. The raw data is uncurated and requires further filtering and processing for analysis, as demonstrated in digital humanities courses at the University of Helsinki. The Reception Reader interface can be used as a web client for exploring the dataset.

**Repository:** Version 1.0.0 is available on Zenodo. DOI: 10.5281/zenodo.7610480.

## (4.3) THE GREAT CHAIN OF DIGITAL TOOLS

The field of digital humanities requires a careful balancing act between research questions, data, and tooling. Given the diverse and ever-evolving nature of research interests, tool building for the field is a continuous journey. One effective strategy is to start by working with a dedicated research group and developing tools from the ground up. By doing so, we start with relevant use cases in mind, before expanding the tools' utility to a broader audience. Perhaps the best example of this is the Old Bailey Online project and decades of producing various kinds of derivative datasets and tools (Hitchcock & Shoemaker, 2006). No research tool operates in a vacuum. The applications are situated in the context of a community of researchers and an ecosystem of digital infrastructure, both of which gain further momentum by being integrated with digital humanities education (Tolonen, Matres, et al., 2018; Tolonen et al., 2020).

The Reception Reader itself has a rich history, born from the collaboration between humanities questions and data science capabilities. Over the years, the team has been developing a broader tooling concept, Octavo, where prototypes were first built to showcase the idea of bringing together distant and close reading (Mäkelä et al., 2020; Tolonen et al., 2017). The idea of visually presenting the overall reception of a work through a single graph was also experimented with using static plotting techniques (Tolonen, Vaara, et al., 2018).

The Reception Reader, therefore, was not planned and developed as a "from beginning to end" solution for a single funded project. Instead, it rises from the shoulders of years of related research projects, and is just another step in the progression of digital humanities tools. When considering the funding of research infrastructure for digital humanities, it is crucial to keep in mind this kind of interconnected progression, and support research-oriented groups and ensure that the work of these groups can be scaled up.

Care and ongoing effort are required to develop the tooling infrastructure and match the tools with research questions and data in order to realise the full potential of a dataset or concept. Earlier interfaces have been developed for text reuse in various digital humanities projects (Gladstone & Cooney, 2020; Vesanto et al., 2017), and the Reception Reader is one next step on this journey, with further advancements to come.

## (4.4) FUTURE DEVELOPMENT

What users can do with the tool is more interesting than its current features, and a main motivation for presenting the information in this article is to gain a further understanding of the users' needs and potential use cases. We aim to gather feedback and incorporate new ideas into the subsequent development of the Reception Reader and other tools. Meanwhile, we already have further improvements in the works, examples are as follows:

**Defragmentation**: We have built a Spark pipeline and run a global clustering process on the underlying datasets to address fragmentation errors and consolidate logical sequences of reuse interrupted by OCR noise, and to remove intermediary reuse links by preserving only the edges connected to the earliest instance of the text segment. This process reduced the 1.9 billion two-way edges down to 373 million defragmented passages, then further down to 49 million "first-source" clusters of passages. We will assess if this consolidated dataset helps provide users with more clarity when exploring reuse patterns.

**Navigation and filtering**: We plan to add labelling and filtering features to help the user "slice and dice" the data points to zoom in on reuse instances of particular interest, while muting the noise from confounding artefacts. To support the use case of exploring multiple works of an author, or multiple editions of a work, navigation and bookmarking features will be added based on the specific research needs of the users. In future iterations, we also want to enable the experience of "walking through" a document to see all the reuses sequentially, with a fast preview of destination pages in the connected documents.

**Representativeness**: Data visualisation maps quantities in a dataset to visual dimensions such as positioning, colour, size, and so on to give the viewer a direct impression of the data. Quantities and relationships can be reflected in aggregate. For example, in the beeswarm chart of the Reception Reader, the density and clustering of connected documents are supposed to offer clues to "which parts of the book were the most discussed and which were not" and show trends over time. The chart aims to visualise the "volume" of reuse in reception. The "true volume", however, is still obscured because of many artefacts, imprints of publishers and advertisement may show up as heavily reused segments, a single document may be commenting on many sections of a source essay, which may then get reprinted, showing up as larger clusters and skewing the perception of reuse visually. In the near future, we want to defragment adjacent reuse instances, deal with confounding artefacts, and refine the charting mechanism to tell a more accurate visual story of the data.

**Infrastructure tooling**: As part of Reception Reader's backend, a number of data services were implemented to support metadata enrichment, offset mapping, and data pre-processing. As a next step, we can extend the tooling and data curation efforts, for example, adding endpoints for retrieving the offset range of a page by page number and for querying for the OCR text of a given page. Currently, the page mapping indices are supplied by Octavo at the document level, further work to enable querying at the page level will make the experience of browsing through different documents smoother. Each time we add to the constellation of data service capabilities, we can do more on the frontend in terms of user interactions that depend on certain types of data queries. It also accelerates subsequent feature development, even across different applications and research projects.

**Comparing editions**: In the eighteenth century, authors and publishers often made significant changes between editions of the same work. The text reuse fragments can be used to compare multiple editions and identify gaps and insertions within them at scale. We can use this information, for example, to trace the evolution of a work over time at the hands of authors, publishers, or censoring authorities, or in the production of a critical edition. By highlighting the connections between an author's full body of work, text reuse opens up new possibilities for the study of individual writers. In future work, we aim to build features to provide scholars with a visual overview of changes through editions, and allow them to consider connections at granular levels, such as a section or chapter rather than a full document.

**Semantic similarity**: In collaboration with the TurkuNLP group, we want to venture into detecting reuse beyond textual overlaps, that is, to find connections not by comparing sequences of characters, but based on meaning throughout various forms of paraphrasing.

**Generalising to other datasets**: The architecture of the setup of the Reception Reader as a web client and its related data services makes it extendable to cover other materials and datasets, especially those that are openly available, which could be anything from newspapers to nineteenth-century Finnish and Swedish books. The tooling can be scaled to accommodate a variety of similar use cases and research interests as long as the original data has been digitised and made available for research use.

## (5) CONCLUSION

The development of effective research tools in digital humanities requires a targeted approach that aligns with specific research needs. The Reception Reader is an example of this approach, serving a specific purpose while leveraging text reuse data that has potential applications in other areas. We want to start with real users in mind, and build a tool they can immediately use to make their research tasks easier. Then, we also have the opportunity to consider who else might benefit from what has been built.

The Reception Reader is partly a product of the infrastructure development in Finland through DARIAH-FI,[5] aimed at facilitating the scaling of tools originating from research projects. In this regard, it also serves as a test case for the idea of building software by researchers for researchers and for bringing a tool initially intended for internal use to a wider audience, so that researchers working on similar topics can benefit from exploring the data and discover new insights to add to their work.

## DATA ACCESSIBILITY STATEMENT

The Reception Reader is a software demo from the HPC-HD project and will be made available for evaluation in early 2023. Since the current use cases refer to ECCO's source data such as document images, follow-up terms of data access will be negotiated with Gale.

---

5   https://www.dariah.fi.

## ACKNOWLEDGEMENTS

We thank the HPC-HD project and in particular Filip Ginter, Michael Mathioudakis, and Rohit Babbar. We also thank the Helsinki Computational History Group (COMHIS) for facilitating an active intellectual environment, the Finnish IT Center for Science (CSC), EuroHPC, and LUMI for providing computing infrastructure, Gale for supporting broader evaluation of the tool, and DARIAH-FI for its patronage of science and scholarship.

## FUNDING INFORMATION

The HPC-HD project is funded by the Academy of Finland under Grants 1333716 and 1347706.

## COMPETING INTERESTS

The authors have no competing interests to declare.

## AUTHOR CONTRIBUTIONS

**David Rosson**: Software; Visualisation; Writing – original draft. **Eetu Mäkelä**: Conceptualisation; Data curation; Resources. **Ville Vaara**: Data curation; Methodology; Resources. **Ananth Mahadevan**: Validation. **Yann Ryan**: Writing – review & editing. **Mikko Tolonen**: Conceptualisation; Supervision; Writing – original draft.

## AUTHOR AFFILIATIONS

**David Rosson**  orcid.org/0000-0001-7956-2070
Department of Digital Humanities, University of Helsinki, Helsinki, Finland
**Eetu Mäkelä**  orcid.org/0000-0002-8366-8414
Department of Digital Humanities, University of Helsinki, Helsinki, Finland
**Ville Vaara**  orcid.org/0000-0001-7924-4355
Department of Digital Humanities, University of Helsinki, Helsinki, Finland
**Ananth Mahadevan**  orcid.org/0000-0001-5401-5716
Department of Computer Science, University of Helsinki, Helsinki, Finland
**Yann Ryan**  orcid.org/0000-0003-1878-4838
Department of Digital Humanities, University of Helsinki, Helsinki, Finland
**Mikko Tolonen**  orcid.org/0000-0003-2892-8911
Department of Digital Humanities, University of Helsinki, Helsinki, Finland